\documentclass[preprint,3p,times,onecolumn,sort&compress]{elsarticle}

\usepackage{amssymb}
\usepackage{epstopdf}

\journal{Journal of Magnetism and Magnetic Materials}

	\usepackage{fancyhdr}
\renewcommand{\thispagestyle}[1]{} 

\begin{document}

\begin{frontmatter}



\title{Thermodynamic properties of a diluted Heisenberg ferromagnet with interaction anisotropy -– magnetocaloric point of view}

\author[label1]{Karol Sza\l{}owski\corref{label0}}
\ead{kszalowski@uni.lodz.pl}
\author[label1]{Tadeusz Balcerzak}
\author[label2]{Andrej Bob\'{a}k}
\address[label1]{Department of Solid State Physics, University of
{\L}\'{o}d\'{z}, ul. Pomorska 149/153, 90-236 {\L}\'{o}d\'{z},
Poland}
\address[label2]{Department of Theoretical Physics and Astrophysics, Faculty of Science, P.J. \v{S}af\'{a}rik University, Park Angelinum 9, 041 54 Ko\v{s}ice, Slovak Republic}
\cortext[label0]{corresponding author}

\begin{abstract}
The thermodynamics of a site-diluted ferromagnetic Heisenberg model for spin $S=1/2$ with interaction anisotropy in spin space is investigated. The study is aimed at presenting the magnetocaloric properties of such a model, including the entropy and temperature changes in magnetization/demagnetization processes, generalized Gr\"uneisen ratio as well as the quantities characterizing the efficiency of magnetic cooling cycles. The results are obtained using Pair Approximation (PA) method and extensively compared with the Molecular Field Approximation (MFA) calculations. The importance of interaction anisotropy and site-dilution is discussed. The inadequacy of the MFA approach (even on the qualitative level) is found for selected quantities, while PA provides the results which are consistent with the experimentally observed behaviour.   
\end{abstract}


\end{frontmatter}



\section{Introduction}

The study of magnetocaloric effect has more than a century-long history, starting from the discovery of Warburg \cite{Warburg1}. Its application in achieving low temperatures by adiabatic demagnetization dates back to the works of Giauque and Debye \cite{Debye1,Giauque1,Giauque2}. At present the effect has gained an immense applicational potential \cite{TishinBook,Szymczak}, providing a hope for an efficient and environmental-friendly refrigeration technique, which stimulates the experimental search for the optimal materials as well as the studies of its thermodynamics \cite{Gschneidner2,Gschneidner3,Pecharsky1,Kloos1,Pecharsky2,Tishin3}. 

The description of the magnetocaloric phenomena depends crucially on the knowledge of the magnetic entropy of the system as a function of the temperature and external magnetic field. In practice, this requires knowledge of complete thermodynamics of the system in question. However, the exact results are still available only for one-dimensional magnetic models which are exactly soluble in external magnetic field (see for example \cite{Trippe}). Other low-dimensional systems also attract attention \cite{Zhitomirsky1,Schmidt1,Honecker1,Honecker2,Pereira1,Honecker3,Ribeiro}. On the other hand, the results for three-dimensional systems appear to be of primary interest and importance. This motivates the studies of magnetocaloric properties of three-dimensional models using the approximate methods for the thermodynamic description. Some studies focus on the magnetocaloric properties in the vicinity of phase transition \cite{Oliveira1,Tishin1,Tishin2,Ranke1,Spichkin2}. Another issue of interest is the existence of some universal relations for characteristics of magnetocaloric effect \cite{Franco1,Franco2,Franco3,Franco7}.

The aim of the present paper is the study of magnetocaloric effect in a model spin-1/2 Heisenberg diluted ferromagnet with interaction anisotropy taken into account. The model makes it possible to discuss the differences which can occur between the quantum Heisenberg and Ising model in the presence of dilution. For this purpose the Pair Approximation (PA) method, developed in our previous works \cite{Balcerzak1,Balcerzak2}, is used. This method incorporates spin-spin correlations into thermodynamic description. The advantage of PA method over the classical molecular field approximation (MFA) is emphasized. In particular, it is shown that some thermodynamic characteristics, to which applying MFA leads to unphysical results, can be successfully coped with by means of PA.

\section{The model and thermodynamic description of the magnetocaloric effect}
The site-diluted spin-1/2 Heisenberg-type ferromagnet with interaction anisotropy is described by the following Hamiltonian:
\begin{eqnarray}
\label{eq1}
\mathcal{H}=-\displaystyle\sum_{\left\langle i,j\right\rangle}^{}{\left[J_{\perp}\left(S^{x}_i S^{x}_j+S^{y}_i S^{y}_j\right)+J_{z}S^{z}_i S^{z}_j\right]\xi_{i}\xi_{j}}\nonumber \\
-H\sum_{i}^{}{S^{z}_{i}\,\xi_{i}}.
\end{eqnarray}
The external magnetic field $H^z$ is introduced by means of the Zeeman term proportional to $H\equiv -g\mu_{\rm B} H^{z}$.  
The condition $0\leq J_{\perp}\leq J_{z}$ is imposed on the exchange integrals between nearest-neighbours for $z$ spin direction and for in-plane spin direction, which implies that the anisotropy is of the Ising type. The particular choice of $J_{\perp}=J_{z}$ corresponds to the isotropic Heisenberg model, while the other limiting case, $J_{\perp}=0$, describes the pure Ising model. 
The site-dilution is conveniently described by means of the occupation-number operators, $\xi_{i} = 0, 1$, defined for each site of the underlying crystalline lattice.
For a random occupation their configurational averages yield $\left\langle \xi_{i}\right\rangle_{r}=p$ and $\left\langle \xi_{i}\xi_{j}\right\rangle_{r}=p^2$, where $p$ is the magnetic atoms concentration.\\

The thermodynamic description of the model is based on the Pair Approximation (PA), which has been discussed in \cite{Balcerzak1,Balcerzak2}. As far as the magnetocaloric effect is concerned, the approach has not been exploited yet. For the purpose of present application it is briefly outlined below. 

The PA belongs to the cluster variational methods. It exploits the idea of single-site and pair density matrices in the cumulant expansion technique up to the second order cumulants.  It is worthwhile to mention that within this technique the MFA is restricted to the first order cumulants only. The method allows the Gibbs energy calculation from which all the thermodynamic properties can be self-consistently obtained. In case of randomly diluted system the Gibbs energy should be averaged over configurations.  The result for the averaged Gibbs potential per one lattice site can be written in the form: 
\begin{equation}
\frac{\left< G \right>_{r}}{N}=\frac{zp^2}{2}\left(G_2-2\frac{zp-1}{zp}G_1\right),
\label{gibbs}
\end{equation}
where $z$ is the coordination number depending on the crystallographic lattice. The single-site and pair Gibbs energies, denoted by $G_1$ and $G_2$, respectively, are given by:
\begin{equation}
    G_1=-k_{\mathrm B} T \ln \left[ 2 \cosh \left(\frac{zp\lambda +H }{2k_{\rm B}T} \right) \right]
\label{}
\end{equation}
and
\begin{eqnarray}
\lefteqn{G_2=-k_{\mathrm B} T \ln \left\{ 2\,\exp\left(\frac{J_{z}}{4k_{\rm B}T}\right)\cosh
    \left[\frac{\left(zp-1\right)\lambda+H}{k_{\rm B}T}\right]\right. }\nonumber\\
    &&+\left.2 \,\exp\left(-\frac{J_{z}}{4k_{\rm B}T}\right)\cosh\left(\frac{J_{\perp}}{2k_{\rm B}T}
    \right) \right\}.
\label{}	
\end{eqnarray}
The value of the variational parameter $\lambda$ follows from the minimum condition for the averaged Gibbs potential $\left<G\right>_r$ and, at the same time, ensures the consistency of magnetizations calculated form single-site and pair density matrices. From this condition, $\lambda$ is determined from the formula:
\begin{equation}
    \!\!\!\!\!\!\!\!\!\!\!\tanh \left(\frac{ zp\lambda +H}{2k_{\rm B}T} \right)=\\
	\frac{\sinh \left[\frac{\left(zp-1\right)\lambda +H}{k_{\rm B}T} \right] }
    {\cosh \left[\frac{\left(zp-1\right)\lambda +H}{k_{\rm B}T} \right]+\exp\left(-\frac{J_{z}}{4k_{\rm B}T}\right) \cosh \left(\frac{J_{\perp}}{2k_{\rm B}T}\right)}.
\label{lambda}	
\end{equation}
The Curie temperature $T_{\rm C}$ can be found when $\lambda \to 0$. As a result of linearization of (\ref{lambda}) one obtains:
\begin{equation}
    \exp \left(\frac{J_{z}}{2 k_{\rm B}T_{\mathrm C}}\right)=\frac{zp}{zp-2} \cosh\left(\frac{J_{\perp}}{2 k_{\rm B}T_{\mathrm C}}\right).
\end{equation}

For the purpose of analyzing the magnetocaloric effect, the knowledge of magnetic entropy as a function of the temperature and external magnetic field, $S\left(T,H\right)$, is crucial. The entropy can be calculated from the Gibbs energy as $S=-\left(\partial \left<G\right>_r / \partial T \right)_{H}$; by the same token the total magnetization of the system $M=-\left(\partial \left<G\right>_r /\partial H\right)_{T}$, as well as the magnetic specific heat $C_{H}=-T\left(\partial^2 \left<G\right>_r /\partial T^2\right)_{H}$ can be obtained. During these calculations, according to eq. (\ref{lambda}), the temperature and field dependence of the parameter $\lambda$ should be taken into account.

The finite entropy change during the isothermal demagnetization process, when the external field changes from $H$ down to 0, is given by the relation $\Delta S_{T}=-\int_{0}^{H}{\left(\partial M/\partial T\right)_{H'}\,dH'}$.

The local sensitivity of the isothermal entropy change $\Delta S_{T}$ to the external field amplitude $H$ can be conveniently quantified by means of the field exponent \cite{Shen1}:
\begin{equation}
\label{eq6}
n=\frac{d\,\ln \Delta S_{T}}{d\,\ln H}=-\frac{H}{\Delta S_{T}}\left(\frac{\partial M}{\partial T}\right)_{H}.
\end{equation}
The value of $n$ means that in the vicinity of a given thermodynamic point $\left(T,H\right)$ the entropy change behaves approximately like $H^{n}$.
 
For the process of magnetization or demagnetization under adiabatic conditions, the temperature change, $\Delta T_{S}$, is often considered. It is obtained form the expression $\Delta T_{S}=-\int_{0}^{H}{\left(T/C_{H}\right)\left(\partial M/\partial T\right)_{H'}\,dH'}.$ During this adiabatic process the magnetization changes accordingly between the external field $H=0$ and $H>0$.

In connection with the adiabatic cooling process, another quantity of interest can be defined, namely 
\begin{equation}
\label{eq10}
\Gamma_{H}=-\frac{1}{T}\left(\frac{\partial S}{\partial H}\right)_{T}/\left(\frac{\partial S}{\partial T}\right)_{H}=-\frac{1}{C_{H}}\left(\frac{\partial M}{\partial T}\right)_{H}.
\end{equation}
This coefficient has an interpretation as the generalized (magnetic) Gr\"uneisen ratio \cite{Zhu1}. $\Gamma_{H}$ is expected to diverge at quantum phase transition point and presents an experimentally measurable quantity \cite{Tokiwa}.

\begin{figure}
\includegraphics[scale=0.55]{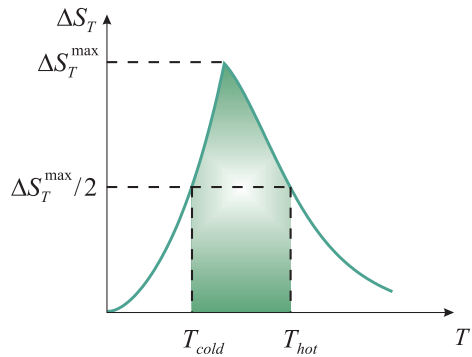}
\caption{Interpretation of a refrigeration capacity (RC) as an area under the dependence of $\Delta S_{T}$ on temperature. The integration range is selected to cover the full-width at half-maximum of the $\Delta S_{T}$ peak.}
\label{FigRCdef}
\end{figure}

A commonly accepted measure for the performance of a substance undergoing  magnetic cooling cycle is the refrigeration capacity (RC), defined as \cite{Gschneidner1} $RC=\int_{T_{cold}}^{T_{hot}}{\Delta S_{T}\left(T\right)\,dT}$,
where the temperatures $T_{cold}$ and $T_{hot}$ are usually selected so as to cover the full-width at half maximum of the entropy change peak (see Fig.~\ref{FigRCdef}).

\begin{figure}
\includegraphics[scale=0.55]{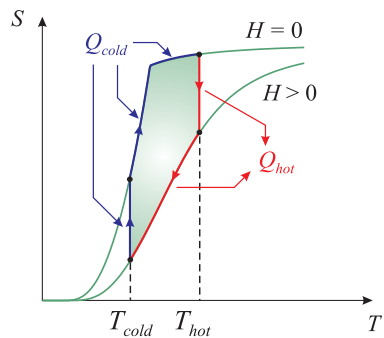}
\caption{Ericsson magnetic cycle working between the external field $H=0$ and $H>0$, presented in the temperature-entropy variables. The heat $Q_{cold}$ is extracted from the cold reservoir during the blue-marked sections, while the heat $Q_{hot}$ is transferred to a hot reservoir during the red-marked sections. }
\label{FigEricsson}
\end{figure}

Among the thermodynamic cycles, important for magnetic refrigeration, the Ericsson cycle is especially worth mentioning \cite{Kitanovski1}. It consists of two isotherms and two processes in constant external field (see Fig.~\ref{FigEricsson}). For such a cooling cycle, the efficiency can be defined as a ratio of the heat extracted from cold reservoir to the work necessary to put in during the cycle, $\eta=Q_{cold}/W=Q_{cold}/\left(Q_{hot}-Q_{cold}\right)$.

It should be emphasized that all the magnetocaloric properties mentioned above can be calculated from one root expression (\ref{gibbs}) for the Gibbs energy.

\section{Numerical results}

\subsection{Test of Pair Approximation method}

\begin{figure}
\includegraphics[scale=0.65]{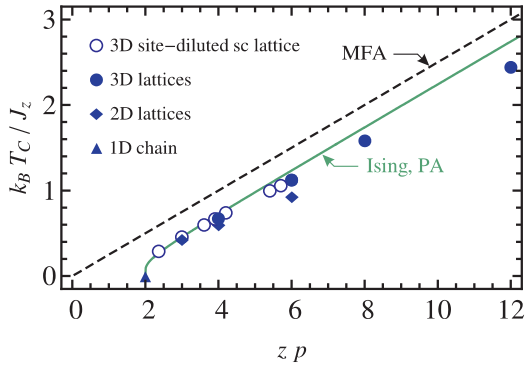}
\caption{The critical temperature of an Ising model for various, one-, two- and three-dimensional lattices, as a function of a coordination number. The solid line is the prediction of PA method, the dashed line is a MFA result. The shaded in symbols show the exact results or precise MC and HTSE estimations (after \cite{Lundow,Butera1,Domb1}). The blank symbols present the MC results for a site-diluted sc lattice (after \cite{Prudnikov,Murtazaev1,Ballesteros1}).}
\label{FigTcPA}
\end{figure}

As a benchmark of the PA method, let us present a comparison of its prediction for critical temperature of the Ising model on various lattices with the exact results (where available) or precise Monte-Carlo (MC) or High Temperature Series Expansion (HTSE) calculations. In the Fig.~\ref{FigTcPA}, the critical temperature is plotted against the  effective coordination number $zp$ for site-diluted model, including the particular case of non-diluted lattices (when $p=1$ and the coordination number is $z$). The plot covers the results for one-, two- and three-dimensional systems. The MFA prediction, $k_{\rm B}T^{MFA}_{c}/J_{z}=zp/4$, is presented by the dashed line, while the PA result, $k_{\rm B}T^{PA}_{c}/J_{z}=1/2\ln\left[zp/\left(zp-2\right)\right]$, is presented by the solid line \cite{Balcerzak1}. The filled symbols represent the accurate (three-dimensional) or exact (one- and two-dimensional) results for various lattices \cite{Lundow,Butera1,Domb1}. The empty symbols represent the MC calculations for site-diluted sc lattice \cite{Prudnikov,Murtazaev1,Ballesteros1}. It is visible that the predictions of PA reproduce satisfactorily the values of critical temperature and the accuracy is remarkable especially for 3D systems with $zp\leq 6$. Also the vanishing of the critical temperature for 1D system (with $z=2$ and $p=1$) is reflected in PA results, but absent in MFA predictions.

\begin{figure}
\includegraphics[scale=0.65]{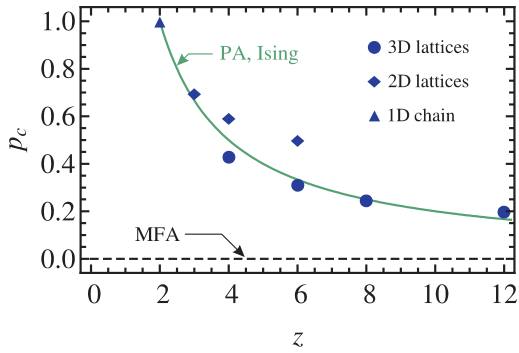}
\caption{The critical concentration for a site-diluted Ising model for various, one-, two- and three-dimensional lattices, as a function of a coordination number. The solid line is the prediction of PA method, the dashed line is a MFA result. The shaded in symbols show the exact results or precise estimations for the site-percolation threshold (after \cite{Newman1,Lorenz1,Lorenz2,Marck1}).}
\label{FigpcPA}
\end{figure}

The analysis can be supplemented by the presentation of the critical concentration for diluted Ising systems, below which the ferromagnetic order vanishes. Such a phenomenon occurs when the concentration drops below the site-percolation threshold. Fig.~\ref{FigpcPA} presents the critical concentration as predicted by PA, i.e. $p_{c}=2/z$, and compared with the site-percolation thresholds, which are known for various lattices \cite{Newman1,Lorenz1,Lorenz2,Marck1}. Again, the agreement is especially striking for 3D systems and the vanishing of magnetic order for 1D chain (described by $p_{c}=1$) is correctly reproduced by PA. It should be noticed that MFA improperly predicts the existence of a ferromagnetic order down to the lowest magnetic component concentrations, i.e., $p_{c}=0$ for all lattices.      

On the basis of the presented results, we can conclude that PA provides a useful and reliable approximate method of constructing the thermodynamic description for the diluted Ising model and is accurate for small $zp$. One also should not overlook the fact that the PA method has successfully been tested for the anisotropic Heisenberg model \cite{Balcerzak2}, including dilution \cite{Balcerzak1}. Therefore, we decided to choose this approach for studies of the magnetocaloric effect. All the further numerical results are presented for the coordination number $z=6$ (3D sc lattice), for which, in the light of previous discussion, the method is expected to be sufficiently accurate. This follows from the consistency of the critical temperature predictions between PA and MC results for crystalline and site-diluted sc lattice (see Fig.~\ref{FigTcPA}), as well as reliable prediction of critical concentration (see Fig.~\ref{FigpcPA})).

\begin{figure}
\includegraphics[scale=0.65]{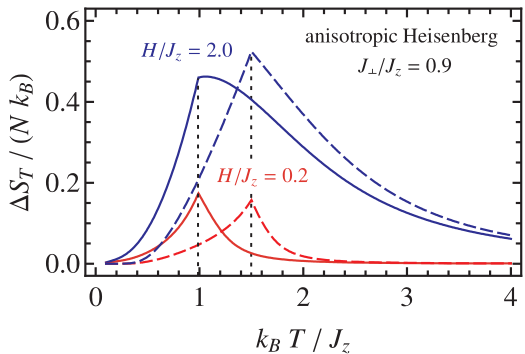}
\caption{Normalized entropy change per lattice site in the isothermal demagnetization process from the external field $H$ down to $H=0$, for the anisotropic Heisenberg model ($J_{z}/J_{\perp}=0.9$), for two values of $H$. The solid lines are the PA results, the dashed lines are MFA results. The position of $T_{\rm C}$ for each case is marked by the vertical dotted line.}
\label{FigDS1}
\end{figure}


\begin{figure}
\includegraphics[scale=0.65]{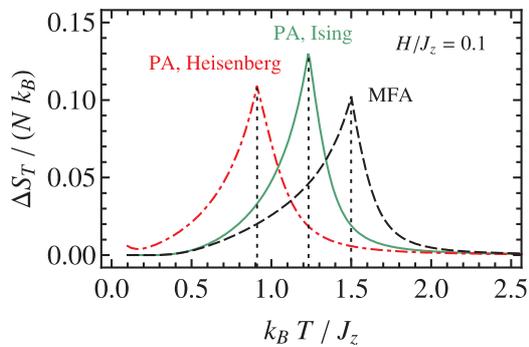}
\caption{Normalized entropy change per lattice site in the isothermal demagnetization process from the external field $H/J_{z}=0.1$ down to $H=0$. The solid line is the PA result for the Ising model, the dashed-dotted line is the PA result for isotropic Heisenberg model, the dashed line is the MFA prediction. The position of $T_{\rm C}$ for each case is marked by the vertical dotted line.}
\label{FigDS2}
\end{figure}

\begin{figure}
\includegraphics[scale=0.65]{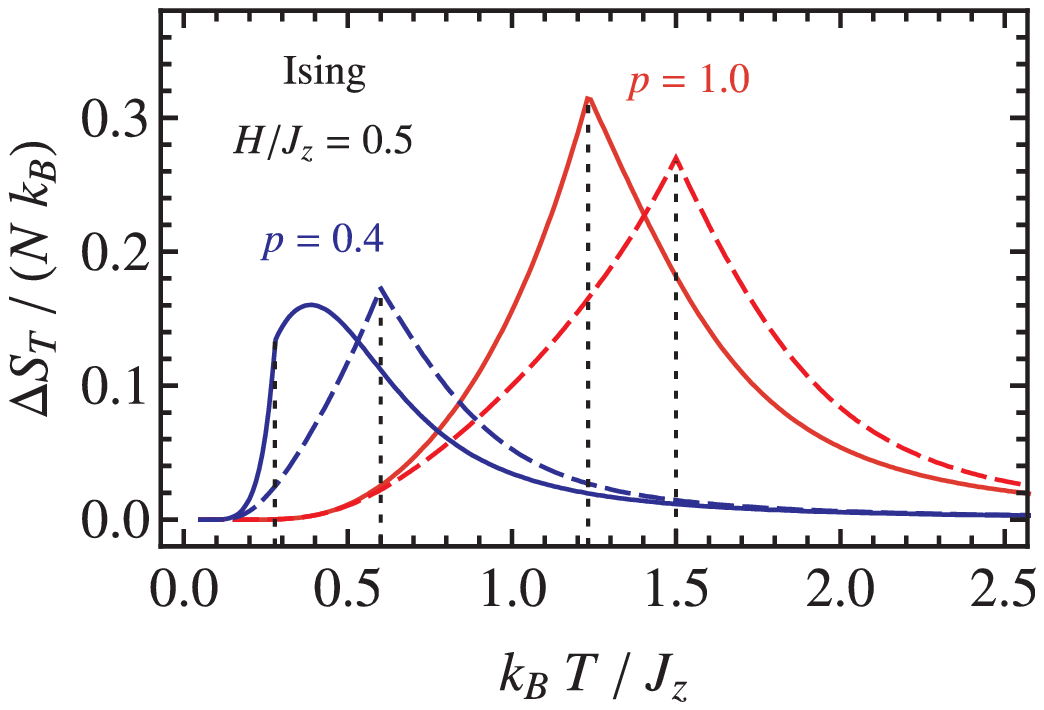}
\caption{Normalized entropy change per lattice site in the isothermal demagnetization process from the external field $H/J_{z}=0.5$ down to $H=0$, for the Ising model, for crystalline and site-diluted case. The solid lines are the PA results, the dashed lines are MFA results. The position of $T_{\rm C}$ for each case is marked by the vertical dotted line.}
\label{FigDSp}
\end{figure}

\subsection{Entropy change in isothermal process}

The isothermal entropy change, $\Delta S_{T}$, when the external field vanishes from $H>0$ to $H=0$, has been studied numerically and some of the results are presented in Figs.~\ref{FigDS1} to \ref{FigDSp}. In Fig.~\ref{FigDS1} $\Delta S_{T}$ is plotted vs. dimensionless temperature $k_{\rm B}T/J_{z}$ for the anisotropy exchange parameter $J_{\perp}/J_{z}=0.9$ and two different fields: $H/J_{z}=0.2$ and $H/J_{z}=2$. The crystalline case is considered. Dashed lines correspond to the MFA while the solid ones represent the PA method. It is seen in Fig.~\ref{FigDS1} that for the low field the maximum of entropy change occurs at the critical temperature and a sharp peak is pronounced. However, for the high field the PA method shows a broadening of the peak from the paramagnetic site, indicating that its shift toward $T>T_{\rm C}$ is possible. A similar behaviour has been discussed in the Ref.~\cite{Franco4}.

In Fig.~\ref{FigDS2} $\Delta S_{T}$ is plotted vs. temperature for the constant field $H/J_{z}=0.1$ and two limiting cases: the crystalline Ising and Heisenberg models. This demonstrates the influence of the anisotropy interaction, which is taken into account by the PA method while the MFA calculation (dashed line), not influenced by the anisotropy, results in both models being indistinguishable. 

The influence of site-dilution on the entropy change in isothermal process of magnetization/demagnetization is depicted in Fig.~\ref{FigDSp}, for the Ising model and for rather strong external field amplitude of $H/J_{z}=0.5$. It is visible that position of the peak ($T_{\rm peak}$) follows the critical temperature change, which decreases with decreasing magnetic component concentration $p$. Also the magnitude of the effect at the maximum is reduced which is connected with the reduction of the magnetization magnitude in a diluted system. It is worth noticing that PA, unlike MFA (dashed line), predicts some widening of the peak when the critical concentration is approached ($p_{c}=1/3$ in case of Fig.~\ref{FigDSp}). 
The widening of the $\Delta S_{T}$ profile for strong dilution regime, as well as some shift of the peak towards $T >T_{\rm C}$, is correlated with the anomalous behaviour (broad paramagnetic maximum) of the magnetic specific heat reported earlier \cite{Balcerzak3} (in the Ising systems).
The lack of coincidence between $T_{\rm C}$ and $T_{\rm peak}$ as a result of dilution is similar to the same phenomenon observed in strong external fields \cite{Franco4}.  The phenomenon seems to appear for the models going beyond MFA, both for the Ising and Heisenberg. In our opinion the effect is worth further investigation.

\begin{figure}
\includegraphics[scale=0.65]{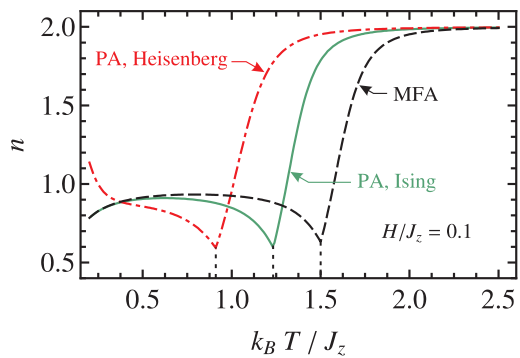}
\caption{The $n$ exponent as a function of the temperature for the external magnetic field $H/J_{z}=0.1$. The solid line is the PA result for the Ising model, the dashed-dotted line is the PA result for isotropic Heisenberg model, the dashed line is the MFA prediction. The position of $T_{\rm C}$ for each case is marked by the vertical dotted line.}
\label{Fign1}
\end{figure}

\begin{figure}
\includegraphics[scale=0.65]{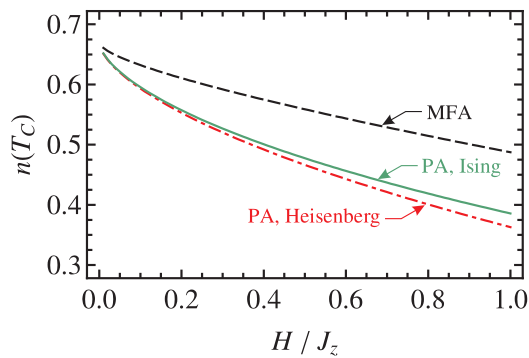}
\caption{The $n$ exponent at the critical temperature, as a function of the normalized external magnetic field. The solid line is the PA result for the Ising model, the dashed-dotted line is the PA result for isotropic Heisenberg model, the dashed line is the MFA prediction.}
\label{Fign2}
\end{figure}

The sensitivity of isothermal entropy change to the magnetic field is quantified by the field exponent $n$ and presented in Figs.~\ref{Fign1} and \ref{Fign2} for the crystalline case. In Fig.~\ref{Fign1} $n$ is plotted vs. temperature in the field $H/J_{z}=0.1$ for the Ising and Heisenberg models. The MFA result (common for both models) is plotted by the dashed line. It can be seen in Fig.~\ref{Fign1} that a sharp minimum of $n$-exponent is observed at the Curie temperature. The value of $n(T_{\rm C})$ at the minimum tends to the limit $n(T_{\rm C})=2/3$, when $H \to 0$. This result is valid both for the Ising and Heisenberg models, which is demonstrated in Fig.~\ref{Fign2} where $n(T_{\rm C})$ is plotted vs. magnetic field. However, when the field increases then $n(T_{\rm C})$ decreases, and this effect is much more pronounced within the PA method than MFA.
This indicates that the dependence of entropy change on field amplitude tends to saturate faster than predicted by MFA, at the same time being weakly dependent on the interaction anisotropy. The behaviour of the index $n$ has been discussed in Refs. \cite{Franco1,Franco2,Franco3,Franco7}, \cite{Franco4}, \cite{Dong1}, \cite{Oesterreicher1}. 

For instance, it has been shown in Refs.\cite{Franco1} and \cite{Franco4} that when the scaling arguments are used the index $n$ for $T=T_{\rm C}$  should not depend on the field. These scaling arguments are also connected with application of the Arrott-Noakes equation of state, which is an approximate semi-empirical formula. It should be mentioned that the best fit of the Arrott-Noakes equation to the experimental data needs several material parameters. However, the results following such approach, for instance, $\Delta S_T$ vs. $H$, reproduce well the corresponding experimental data. On the other hand, in our approach the equation of state follows from the thermodynamic relationship $M=-\left(\partial \left<G\right>_r /\partial H\right)_{T}$ and the only fitting parameters are the exchange integrals in the Hamiltonian. Thus, our method seems more self-consistent from the thermodynamic point of view. However, the price for it is a less precise agreement with the experimental data, especially in the vicinity of the critical point. Let us emphasize that the Arrot-Noakes equation of state seems particularly useful as a tool for empirical description of rather complex systems. 

The dependence of magnetic entropy change on the external field has been recently modeled in the Ref.~\cite{Lyubina1}, where it has been emphasized that the proportionality of $\Delta S_{T}$ to $H^{2/3}$ at critical temperature has a limited range of validity.


\subsection{Temperature change in adiabatic process}

\begin{figure}
\includegraphics[scale=0.65]{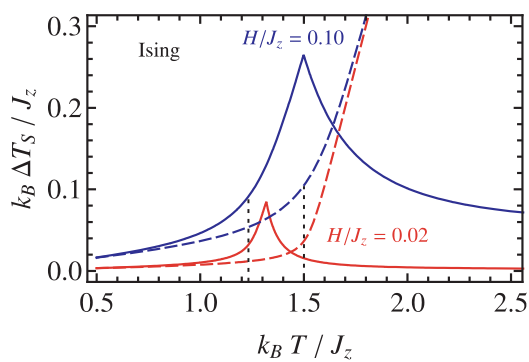}
\caption{Normalized temperature change in the adiabatic magnetization process from the external field $H=0$ up to $H$ for the Ising model, for two values of $H$. The solid lines are the PA results, the dashed lines are MFA results. The positions of $T_{\rm C}$ are marked by the vertical dotted lines.}
\label{FigDT1}
\end{figure}

\begin{figure}
\includegraphics[scale=0.65]{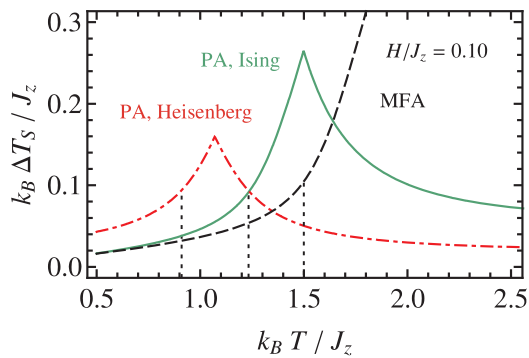}
\caption{Normalized temperature change in the adiabatic magnetization process from the external field $H=0$ up to $H/J_{z}=0.1$ for the Ising model, for two values of $H$. The solid line is the PA result for the Ising model, the dashed-dotted line is the PA result for isotropic Heisenberg model, the dashed line is the MFA prediction. The positions of $T_{\rm C}$ are marked by the vertical dotted lines.}
\label{FigDT2}
\end{figure}

The temperature change, $\Delta T_{S}$, has been studied in relation to the process of adiabatic magnetization and the results are presented in Figs.~\ref{FigDT1} and \ref{FigDT2}. The data are plotted vs. dimensionless temperature, $k_{\rm B}T/J_{z}$, for which the field $H>0$ was applied. In Fig.~\ref{FigDT1} the results for the crystalline Ising model in two magnetic fields: $H/J_{z}=0.02$ and $H/J_{z}=0.10$ are presented. The dashed lines correspond to the MFA method. Contrary to the PA, where $\Delta T_{S}$
curves show some characteristic peaks, the MFA method gives unphysical divergence of $\Delta T_{S}$ when the temperature goes to infinity. Such behaviour is connected with the fact that in MFA the entropy is constant above the Curie temperature. The divergence is not observed experimentally. Yet the peaks (increasing with the field magnitude) have been measured in many ferromagnetic systems (see for example \cite{Franco6}, \cite{Dankov1}). Thus, the PA results for $\Delta T_{S}$ are qualitatively more correct than those of MFA.

In Fig.~\ref{FigDT2} the comparison of the Ising and Heisenberg models is made in the constant magnetic field $H/J_{z}=0.10$. Again, the MFA results are indicated by the dashed line. It can be noticed that for the Heisenberg model the peak is less pronounced than for the Ising one and is shifted toward lower temperature. In both figures \ref{FigDT1} and \ref{FigDT2}, in the low temperature region, the results of the MFA and PA methods converge for the Ising case. A difference with the Heisenberg case, as seen in Fig.~\ref{FigDT2}, originates from the residual entropy, which exists in the isotropic quantum system when $T \to 0$ \cite{Balcerzak2}.

\subsection{Generalized Gr\"uneisen ratio}

\begin{figure}
\includegraphics[scale=0.65]{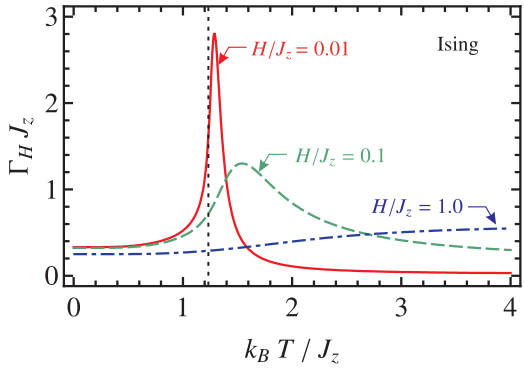}
\caption{Generalized Gr\"uneisen ratio as a function of normalized temperature, for the Ising model. The solid line is for the external field $H/J_{z}=0.01$, the dashed one for $H/J_{z}=0.1$, the dashed-dotted one for $H/J_{z}=1.0$. The position of $T_{\rm C}$ is marked by the vertical dotted line.}
\label{FigGamma1}
\end{figure}

\begin{figure}
\includegraphics[scale=0.65]{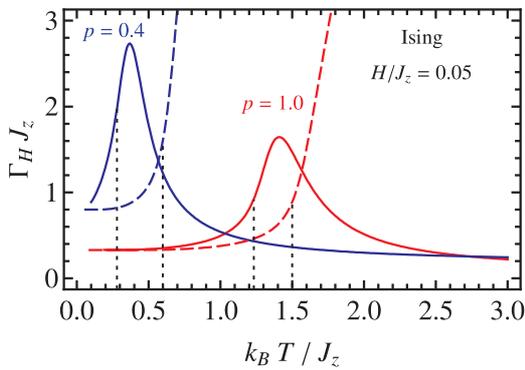}
\caption{Generalized Gr\"uneisen ratio as a function of normalized temperature, for the Ising model in the external field $H/J_{z}=0.05$, for crystalline and site-diluted case. The solid lines are the PA results, the dashed lines are MFA results. The positions of $T_{\rm C}$ are marked by the vertical dotted lines.}
\label{FigGammap}
\end{figure}

The Gr\"uneisen ratio for magnetic systems has been defined by eq. 
(\ref{eq10}) and its calculations vs. temperature are presented in Figs.~\ref{FigGamma1} and \ref{FigGammap}. In Fig.~\ref{FigGamma1} the results for the crystalline Ising model is plotted in PA for several magnetic field values: $H/J_{z}=0.01$; 0.1 and 1. It is seen that for low fields a peak occurs, which is connected with the temperature derivative of magnetization. For larger fields the peak becomes broadened and shifted toward higher temperatures.

\begin{figure}
\includegraphics[scale=0.65]{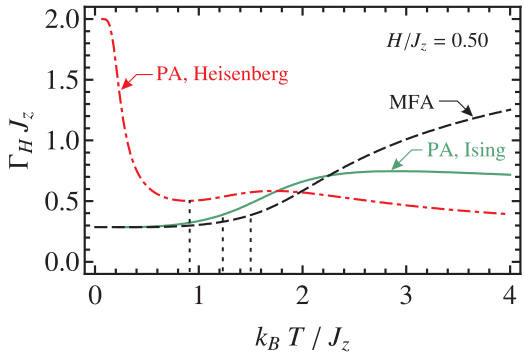}
\caption{Generalized Gr\"uneisen ratio as a function of normalized temperature, for the external field $H/J_{z}=0.50$. The solid line is the PA result for the Ising model, the dashed-dotted line is the PA result for isotropic Heisenberg model, the dashed line is the MFA prediction. The positions of $T_{\rm C}$ are marked by the vertical dotted lines.}
\label{FigGamma2}
\end{figure}

In Fig.~\ref{FigGamma2}, for a relatively large field, $H/J_{z}=0.5$, a comparison between the Heisenberg and Ising models is made. The dashed line corresponds to the MFA calculation. We see that for very low temperatures the MFA result agrees well with that of PA for the Ising model. However, for the Heisenberg model an increase of $\Gamma_{H}$ is predicted, which results from the interplay between $C_{H}$ and $\left(\partial M/\partial T\right)_{H}$ behaviour in a quantum system. We think that increase of $\Gamma_{H}$-ratio when $T \to 0$ has not been reported yet. 
An explanation for such increase can easily be made based on the spin-wave theory, where $\left(\partial M/\partial T\right)_{H} \propto T^{1/2}$ on the one hand, and $C_{H} \propto T^{3/2}$ on the other hand. Thus, at low temperatures and without external field, the $\Gamma_{H}$-ratio given by eq. (\ref{eq10}) should diverge like $\propto 1/T$. However, in the presence of external field the Zeeman term modifies the spin-wave energy, and the specific heat behaves like $C_{H}=aT^{1/2}+bT^{3/2}$. This, in consequence, leads to the finite value of $\Gamma_{H}$ in the limit $T \to 0$, in accordance with Fig.~\ref{FigGamma2}.
We have checked that for intermediate anisotropy parameters, $0<J_{\perp}/J_{z}<1$, the $\Gamma_{H}$-curves lie between those for the Ising and Heisenberg models.

The effect of dilution on the $\Gamma_{H}$-ratio  is demonstrated in Fig.~\ref{FigGammap} for the Ising model and $H/J_{z}=0.05$. Some peaks characteristic of the PA method occur and they are shifted toward lower temperatures while the concentration of magnetic atoms, $p$, decreases. The dashed lines correspond to the MFA results. It is seen that although the results of the PA and MFA methods converge in the low temperatures region, they are completely different for high temperatures. The divergence of the dashed curves presented in Fig.~\ref{FigGammap} is evidently improper and is connected with the fact that above the Curie temperature the magnetic specific heat vanishes in the MFA method. Another interesting feature seen in Fig.~\ref{FigGammap} is that for very high temperatures the PA results become independent of the concentration, approaching the same limit. This effect is worthy of further studies and possible experimental verification.

\subsection{Efficiency of magnetic refrigeration}

\begin{figure}
\includegraphics[scale=0.65]{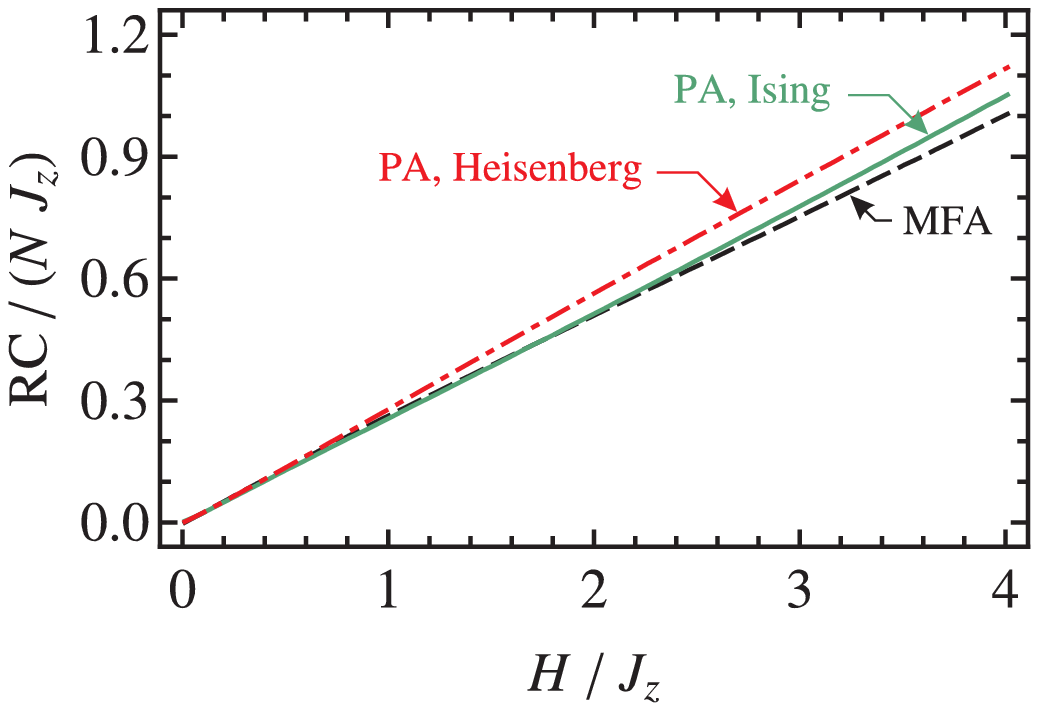}
\caption{Normalized refrigeration capacity per lattice site as a function of the normalized external field. The solid line is the PA result for the Ising model, the dashed-dotted line is the PA result for isotropic Heisenberg model, the dashed line is the MFA prediction.}
\label{FigRC2}
\end{figure}

The values of RC, quantifying the performance of the system in refrigeration (Ericsson) cycle, appear to change almost linearly with the amplitude of external field, according to both the MFA and PA predictions. This property is illustrated in Fig.~\ref{FigRC2} for the Ising and Heisenberg models, together with the MFA results which are denoted by the dashed line. It is seen in Fig.~\ref{FigRC2} that the influence of interaction anisotropy is very weak for this characteristic and becomes observable only when a strong field is used. The linear or near-linear RC dependence on external field is experimentally observed, for example in the results of Refs.~\cite{Franco2}, \cite{Arora1,Arumugam1,Franco5,Bingham1}. 

It should be mentioned that in the recent paper \cite{Franco1} the non-linear, power dependence of RC vs. $H$ is predicted, based on the scaling arguments. That result is again connected with application of the Arrott-Noakes equation of state for calculations of RC.  In order to explain the discrepancy with our almost linear result one should note that the calculation of RC is based on calculation of $\Delta S_T$. Thus, the problem becomes essentially similar to that discussed in the context of the Fig.~\ref{Fign2}, i.e., it reduces to the choice of the equation of state.

\begin{figure}
\includegraphics[scale=0.65]{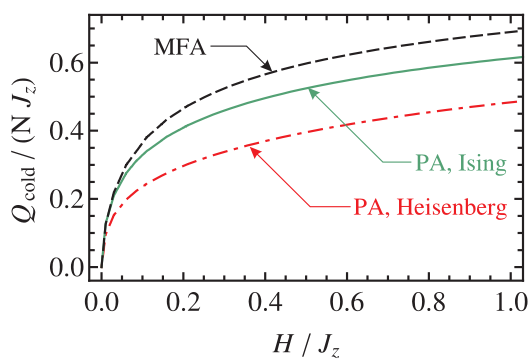}
\caption{Normalized heat per lattice site extracted from the cold reservoir in one Ericsson cycle working between the external fields $H=0$ and $H$. The solid line is the PA result for the Ising model, the dashed-dotted line is the PA result for isotropic Heisenberg model, the dashed line is the MFA prediction.}
\label{FigQcold}
\end{figure}

Another important quantity describing the real effectiveness of the refrigeration process is the amount of heat $Q_{cold}$ extracted from the cold reservoir in one cycle. As far as the Ericsson cycle is concerned the dependence of $Q_{cold}$ on the field amplitude used in the cooling cycle is presented in Fig.~\ref{FigQcold}. The initial fast increase of the curves tends to saturate for larger fields. It is observable that MFA markedly overestimates the value of exchanged heat by a significant value when compared to the PA calculations. The influence of interaction anisotropy is also clearly pronounced here, as for the Ising model the values of $Q_{cold}$ are noticeably higher than for the isotropic Heisenberg one.

\begin{figure}
\includegraphics[scale=0.65]{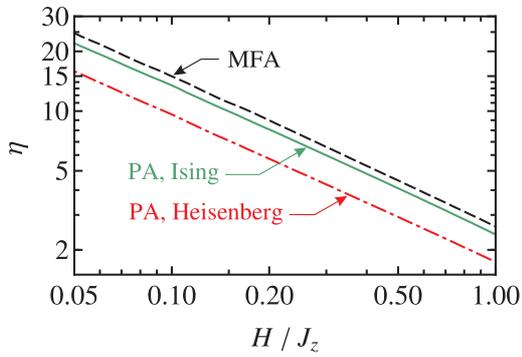}
\caption{Efficiency of the Ericsson cycle working between the external fields $H=0$ and $H$. The solid line is the PA result for the Ising model, the dashed-dotted line is the PA result for isotropic Heisenberg model, the dashed line is the MFA prediction.}
\label{FigEtaEricsson}
\end{figure}

The above results should be compared with calculations of the cooling cycle efficiency, $\eta_{Ericsson}$, which is presented in Fig.~\ref{FigEtaEricsson}. Here it can be observed that using larger external field amplitude the efficiency of the cycle rapidly decreases. Again, the MFA method overestimates the efficiency $\eta_{Ericsson}$; however, there is only a slight difference between MFA and PA results for the Ising model. Yet the PA predictions for the isotropic Heisenberg model are much lower than that of MFA.

It can be deduced from Figs.\ref{FigQcold} and \ref{FigEtaEricsson} that an increase of the field amplitude limits severely the efficiency on the one hand, but it increases the absolute amount of heat extracted during one cycle on the other hand. Thus, in order to design a well-performing cycle, a compromise between both tendencies must be found, for high efficiency with weak field would require fast repetition of the cycle to gain assumed cooling power (or large amount of magnetic working substance).

\begin{figure}
\includegraphics[scale=0.65]{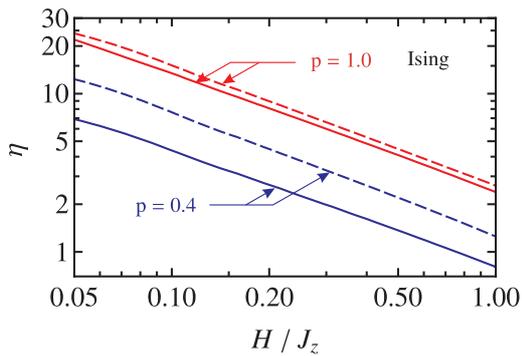}
\caption{Efficiency of the Ericsson cycle working between the external fields $H=0$ and $H$, for the Ising model, for crystalline and site-diluted case. The solid lines are the PA results, the dashed lines are MFA results.}
\label{FigEtaEricsson2}
\end{figure}

The effect of site dilution on $\eta_{Ericsson}$ is presented in Fig.~\ref{FigEtaEricsson2} for the Ising model. By comparison with Fig.~\ref{FigEtaEricsson} it is seen that dilution decreases the cycle efficiency and, at the same time, the difference between the PA and MFA methods becomes more visible.

\section{Final remarks}

We performed extensive calculations to illustrate the magnetocaloric properties of the anisotropic Heisenberg model on the s.c. lattice. 
The initial comparison of the predictions of PA and MFA for the critical temperature as well as the critical concentration of the Ising model has satisfactorily tested the applicability of our method. The use of PA allows us to study the influence of interaction anisotropy on the thermodynamic properties, which is especially noteworthy considering that MFA remains completely insensitive to this parameter. Moreover, the predictions of PA referring to the influence of site-dilution are much more reliable that that of MFA. Such improvement is caused by taking the spin-pair correlations into account, which implies that the internal energy, as well as magnetic entropy, are better calculated.

We have shown that MFA may lead to incorrect results even at the qualitative level, with regard to the generalized Gr\"uneisen ratio and the temperature difference in adiabatic magnetization/demagnetization process. Unlike MFA, PA provides the results which are mostly consistent with the experimentally observed behaviour of the ferromagnetic systems. 
The predictions of PA and MFA differ also at the quantitative level. As an example, MFA overestimates the efficiency of the Ericsson cycle by a significant value or predicts much slower decrease of the $n$ index with the amplitude of the external field. On the basis of all those comparisons one can generally conclude that the PA method, being more accurate than MFA, can be recommended from the methodological point of view for the description of a self-consistent thermodynamics.

Let us emphasize that there are numerous efforts aimed at finding universal relations for the magnetocaloric effect by using the scaling approach (see \cite{Franco2,Dong1,Franco3,Franco1,Franco5}). These models, however, are based on the phenomenological equations of state like the Arrott-Noakes equation, or expansions of the free-energy within the Landau model of phase transitions, with the coefficients known from the experiment \cite{Amaral1,Amaral2,Shen2}, successful at describing empirically complex magnetic systems. Our model, by contrast, is constructed on fully microscopic grounds. It starts from the Hamiltonian level, and hence the only parameters entering the calculations are the exchange integrals.

Further developments of the presented approach might include extension of the studies to the models with XY-type interaction anisotropy and systems with antiferromagnetic interactions, especially those in which quantum critical point behaviour can be expected. Another issue are the investigations of low-dimensional magnets, for which the PA method is also applicable. For more general thermodynamics, the lattice and electronic contribution to the total entropy of magnetic systems would be worth consideration.

\bibliographystyle{elsarticle-num}







\end{document}